\documentclass[%
preprint,
amsmath,amssymb,
aip,
]{revtex4-2}
\usepackage{graphicx}
\usepackage{dcolumn}
\usepackage{bm}
\usepackage{hyperref}
\usepackage{color}
\usepackage{url}
\usepackage[english]{babel}
\graphicspath{{../}}

\usepackage{subfiles} 

\begin{document}

\title{Multichannel highly sensitive diamond quantum magnetometer}

\author{A. Yoshimura}
\affiliation{Department of Electrical and Electronic Engineering, Institute of Science Tokyo, Meguro, Tokyo 152-8550, Japan}

\author{A. Kanamoto}
\affiliation{Department of Electrical and Electronic Engineering, Institute of Science Tokyo, Meguro, Tokyo 152-8550, Japan}

\author{N. Sekiguchi}
\email[]{sekiguchi.n.ac@m.titech.ac.jp}
\affiliation{Department of Electrical and Electronic Engineering, Institute of Science Tokyo, Meguro, Tokyo 152-8550, Japan}

\author{C. Shinei}
\affiliation{Research Center for Electronic and Optical Materials, National Institute for Materials Science, Tsukuba, Ibaraki 305-0044, Japan}
\affiliation{Department of Applied Physics, University of Tsukuba, Tsukuba, Ibaraki 305-8571, Japan}

\author{M. Miyakawa}
\affiliation{Research Center for Materials Nanoarchitectonics, National Institute for Materials Science, Tsukuba, Ibaraki 305-0044, Japan}

\author{T. Taniguchi}
\affiliation{Research Center for Materials Nanoarchitectonics, National Institute for Materials Science, Tsukuba, Ibaraki 305-0044, Japan}

\author{T. Teraji}
\affiliation{Research Center for Electronic and Optical Materials, National Institute for Materials Science, Tsukuba, Ibaraki 305-0044, Japan}

\author{H. Abe}
\affiliation{Takasaki Institute for Advanced Quantum Science, National Institutes for Quantum Science and Technology, Takasaki, Gunma 370-1292, Japan}

\author{S. Onoda}
\affiliation{Takasaki Institute for Advanced Quantum Science, National Institutes for Quantum Science and Technology, Takasaki, Gunma 370-1292, Japan}

\author{T. Ohshima}
\affiliation{Takasaki Institute for Advanced Quantum Science, National Institutes for Quantum Science and Technology, Takasaki, Gunma 370-1292, Japan}
\affiliation{Department of Materials Science, Tohoku University, Sendai, Miyagi 980-8579, Japan}

\author{T. Iwasaki}
\affiliation{Department of Electrical and Electronic Engineering, Institute of Science Tokyo, Meguro, Tokyo 152-8550, Japan}

\author{M. Hatano}
\affiliation{Department of Electrical and Electronic Engineering, Institute of Science Tokyo, Meguro, Tokyo 152-8550, Japan}

\date{\today}

\begin{abstract}
We demonstrate a highly sensitive real-time magnetometry method at two measurement points.
This magnetometry method is based on the frequency-division multiplexing of continuous-wave optically detected magnetic resonance.
We use two ensembles of nitrogen--vacancy (NV) centers separated by 3.6~mm to measure a magnetic field.
A different bias field is applied to the two NV ensembles to resolve the resonance peak for each ensemble in the frequency space and enables the multiplexed magnetometry at the two points.
The sensitivities achieved at the measurement points are $21~\mathrm{pT/\sqrt{Hz}}$ and $22~\mathrm{pT/\sqrt{Hz}}$.
The proposed magnetometry method can be expanded to include more measurement points and shorter spacing.
The capability of real-time measurement at numerous points with short spacing and high sensitivity is beneficial for various applications, including biomagnetic sensing, geophysical research, and material science.
\end{abstract}

\maketitle

\section{Introduction}
The measurement of a magnetic field plays a key role in multiple domains, ranging from fundamental physics to material science and biology \cite{Saf18, Chr24, Asl23}.
Diamond quantum magnetometers (DQMs), which use nitrogen-vacancy (NV) centers in diamonds, have attracted considerable attention because they can be operated with high sensitivity and a high spatial resolution at room temperature, even in strong magnetic field environments \cite{Chr24, Asl23, Fu20, Bar20}.
In addition, DQMs can be self-calibrated because their operating principle is based on fundamental physical constants.

One promising application of DQMs is biomagnetic sensing, in which high-sensitivity magnetometry with a high spatial resolution at small measurement distances is desirable \cite{Asl23, Ham93, Kor16, Ara22}.
Room-temperature operation with high sensitivity reduces the standoff distance from a measurement object. In contrast, other ultrasensitive magnetometers, including superconducting quantum interference devices and atomic magnetometers, are operated at low or high temperatures and require an insulation layer \cite{Asl23, Kor16}.
A short standoff distance captures the fine spatial distribution of a biomagnetic field.
Furthermore, it increases the signal amplitude and makes sensing robust against environmental noise.
The distribution should be measured as finely as possible when estimating a field source, such as the intracellular current, by solving an inverse problem.
This requires the real-time measurement of the distribution; that is, the magnetic fields at multiple points should be simultaneously measured.
The required spacing between measurement points is less than or approximately equal to the standoff distance \cite{Ham93, Sek25}.
In addition, a weak biomagnetic field requires high sensitivity.

However, certain essential components of high-sensitivity DQMs prevent the measurement of the real-time spatial distribution of a field with a spacing smaller than approximately 10 mm.
A major factor for increasing the spacing is the large size of the fluorescence collection optics.
In most DQMs with high sensitivities, the fluorescence intensity from NV centers is measured to read out the spin states \cite{Bar16, Ara22, Gra23, Sch18, Wol15, Zha21, Shi22, Bar20, Bar24, Sek24, Sek25}.
The fluorescence collection efficiency affects the photon shot-noise-limited sensitivity at a scale of $-1/2$ power \cite{Bar20}.
Therefore, the collection optics become large to achieve high sensitivity \cite{Le12}.
Another major factor for increasing the spacing is the difficulty in confining a microwave (MW) magnetic field at approximately 3~GHz, which drives the magnetic resonance of NV center spins, to the millimeter scale.
The application of the MW field to multiple DQMs may cause interference and disturb the measurement.
To utilize DQMs, a technique must be developed for simultaneously performing highly sensitive magnetometry at multiple points with short spacing.
Although gradiometers that use the opposite arrangement of single-sided magnetometers have demonstrated real-time measurements with high sensitivity at two points with millimeter-scale spacing \cite{Zha23}, increasing the number of measurement points and placing the magnetometers close to the measurement object are challenging.
Infrared-absorption-based DQMs, which do not require fluorescence collection optics, have demonstrated either high sensitivity \cite{Cha17, Tay25} or real-time measurement at multiple points \cite{Bop25}; however, achieving both simultaneously is difficult.

In this study, we develop a DQM to measure the magnetic fields at two points with a spacing of 3.6~mm simultaneously as the first step toward high-sensitivity field mapping at the millimeter scale.
Sensitivities of $21~\mathrm{pT/\sqrt{Hz}}$ and $22~\mathrm{pT/\sqrt{Hz}}$ at each point are achieved in a bandwidth range of 25--300~Hz.
We employ frequency-division multiplexing to perform field measurements at multiple points.
The magnetic resonance frequencies at the two positions are separated by applying different bias magnetic fields.
By driving these resonances with frequency modulations at different frequencies, the magnetic field at each position is encoded on the frequency components of a time-domain signal.
The demodulation of the frequency components yields the magnetic field at the two points.
The spacing between the measurement points is limited by the geometry of the magnetic coils and the spot size of the excitation laser. The spacing can be decreased to the submillimeter scale in the future.

\section{Measurement procedure}

\begin{figure}[tb]
\centering
\includegraphics[scale=1]{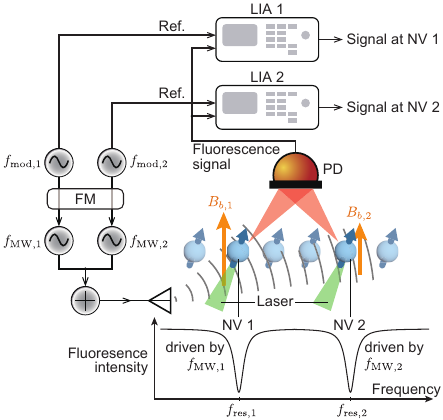}
\caption{\label{fig:procedure}
    Schematic of simultaneous multipoint measurement of a magnetic field using a single magnetometry module.
    The balls with arrows represent NV centers at different positions. 
    Multiple ensembles under different bias magnetic fields $B_{b, i}$ are excited by laser beams.
    The laser-induced fluorescence from both ensembles is detected using a single detector.
    PD: photodiode;
    FM: frequency modulation;
    LIA: lock-in amplifier.
}
\end{figure}

We developed a procedure for simultaneously measuring magnetic fields at different positions using a single magnetometry module based on optically detected magnetic resonance (ODMR) and frequency-division multiplexing.
Fig.~\ref{fig:procedure} shows an overview of the procedure, in which a two-point measurement is illustrated as an example.
We used the ensembles ($\mathrm{NV}_{i}$, $i=1, 2$) of NV centers at different positions and excited them using green laser beams.
The laser-induced fluorescence from the ensembles was detected using a single photodetector.
The resonance frequencies $f_{\mathrm{res},i}$ of the magnetic transitions of the ensembles were varied by applying different bias magnetic fields $B_{\mathrm{b},i}$.
MW fields with different frequencies $f_{\mathrm{MW},i}$ drove magnetic transitions at each ensemble with the frequency modulations at $f_{\mathrm{mod},i}$.
The frequency modulations provided the frequency components at $f_{\mathrm{mod},i}$ in the fluorescence signal, and they reflected the frequency-modulated magnetic resonance signal at each position.
Therefore, the change in the magnetic field at each position was measured in real time by demodulating the fluorescence signal at $f_{\mathrm{mod},i}$.

\section{Experimental setup}
\label{sec: setup}

\begin{figure}[tb]
\centering
\includegraphics[scale=1]{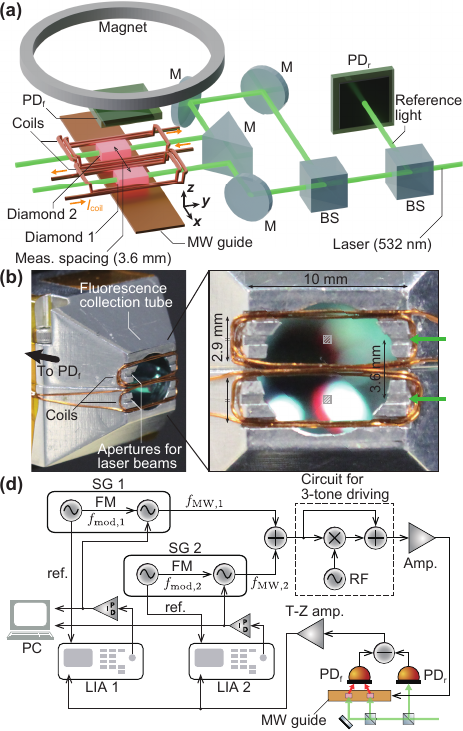}
\caption{\label{fig:opt_setup}%
Experimental setup. 
(a) Schematic of optical setup and coils surrounding the diamonds. 
BS: beam splitter; M: mirror.
(b) Photographs of the coils wound around a fluorescence collection tube with an elliptical reflective inner surface.
The hatched regions indicate the locations of the diamonds.
(c) Circuit diagram. SG: signal generator; RF: radio frequency; T-Z amp.: transimpedance amplifier; PID: proportional--integral--derivative regulator.
}
\end{figure}
We created an experimental setup to measure the magnetic fields at two positions with a spacing of 3.6 mm.
This spacing was based on the requirement for measuring the spatial distribution of the brain magnetic field in small animals such as rats \cite{Sek25, Pax98}.
Fig.~\ref{fig:opt_setup}(a) shows the experimental setup.
We placed two diamonds with surface orientations of (111) at a distance of 3.6 mm.
The diamonds were split from one diamond crystal that was synthesized using a high-pressure–high-temperature (HPHT) method, in which the nitrogen concentration was controlled using a metal solvent \cite{Miy22}.
The concentration of substitutional nitrogen ($\mathrm{N_s^0}$) was estimated to be 1.8~ppm using electron-spin resonance (ESR).
We reduced the $^{13}$C concentration to the order of 50~ppm using a $^{12}$C-enriched carbon source for the HPHT synthesis.
The area and thickness of the diamonds were $1.2~\mathrm{mm}\times0.8~\mathrm{mm}$ and 0.3~mm, respectively.
Vacancies were created in the diamonds using electron beam irradiation.
The energy and total fluence of the electron beam were 2~MeV and $1\times10^{18}~\mathrm{cm}^{-2}$, respectively.
NV centers were produced by annealing the diamonds at 1000~${}^\circ$C for 2 h under vacuum.
The concentrations of residual $\mathrm{N_s^0}$ and negatively-charged NV centers were estimated to be 0.3~ppm and 0.2~ppm, respectively, using ESR.
The dephasing time $T_2^{\ast}$ was estimated to be $T_2^{*}\sim4.6~\mathrm{\mu s}$ based on independent measurements.
Each diamond was irradiated using an excitation laser beam at 532 nm.
The beam spot size was $100~\mathrm{\mu m}$ ($1/e^2$-diameter).
The intensities of the two beams were approximately 1.1 W each.

\begin{figure}[tb]
\centering
\includegraphics{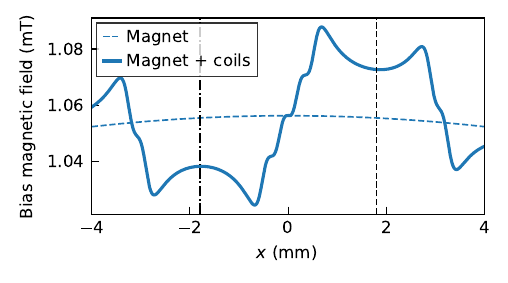}
\caption{\label{fig:mag_setup}
    Calculated bias magnetic field profile along the $x$-axis for $I_{\mathrm{coil}}=25~\mathrm{mA}$.
    The dashed and dash-dotted vertical lines indicate the positions of diamonds 1 and 2, respectively.
}
\end{figure}

We applied a bias field along the $z$-axis $\parallel (111)$ using a permanent magnet and coils that surrounded the diamonds, as shown in Fig.\ref{fig:opt_setup}(a) and (b).
The magnet applied a common bias field of approximately 1~mT, which was modified by the coils depending on their positions.
The magnetic field generated by the coils had the same direction as the common field at diamond 1 and the opposite direction at diamond 2.
The coils were connected in series such that the currents $I_{\mathrm{coil}}$ were identical.
Figure \ref{fig:mag_setup} shows the calculated bias field for $I_{\mathrm{coil}}=25~\mathrm{mA}$ as a function of the $x$ position with respect to the midpoint of the diamonds.
The dashed curve represents the common field generated by the magnet, and the solid curve represents the total bias field generated by the magnet and coils.
The center positions of diamonds 1 and 2 are indicated by the dashed and dash-dotted vertical lines, respectively.
We designed the coils such that the diamonds were located at the extrema of the bias field profile to suppress field inhomogeneities within the excitation beams, thereby reducing the broadening of the ODMR peak.
The coil current $I_{\mathrm{coil}}$ was monitored by measuring the voltage across a $1~\Omega $ shunt resistor using an instrumentation amplifier.
The ratio of the magnetic field to current was calibrated by measuring the Zeeman splitting between the ground states $|1\rangle$ and $|-1\rangle$ as a function of $I_{\mathrm{coil}}$ to be $0.60\pm0.02~\mathrm{\mu T/mA}$ and $-0.65\pm0.02~\mathrm{\mu T/mA}$ for diamonds 1 and 2, respectively.

We detected the laser-induced fluorescence from the two diamonds using a single photodiode.
The fluorescence detection was similar to that in our previous study\cite{Sek24, Sek25}.
We placed hemispherical lenses on each diamond and used a fluorescence collection tube with an elliptical reflective inner surface.
The scattered laser light was filtered out using a long-pass filter with a cut-on wavelength of 650 nm.
We used the balanced detection technique to reduce the intensity noise of the excitation laser light by splitting off a portion of the light as a reference light, as shown in Fig.~\ref{fig:opt_setup}(a).

We applied an MW field using a 0.1-mm-thick copper foil as the MW guide.
The MW circuit is illustrated in Fig.~\ref{fig:opt_setup}(c).
The driving MW field consisted of two components with different center frequencies, $f_{\mathrm{MW, 1}}$ and $f_{\mathrm{MW,2}}$, and different modulation frequencies, $f_{\mathrm{mod, 1}}$ and $f_{\mathrm{mod, 2}}$. 
We used $f_{\mathrm{mod, 1}}=7.5~\mathrm{kHz}$ and $f_{\mathrm{mod, 2}}=10~\mathrm{kHz}$.
We simultaneously drove the three transitions associated with the hyperfine states in the $|0\rangle\leftrightarrow|-1\rangle$ electron spin transition to improve the contrast of ODMR peaks \cite{Bar16}.

This was achieved by mixing the MWs with a radio-frequency (RF) field at a hyperfine splitting of 2.16~MHz and combining the mixed MWs with bypassed MWs [see Fig. ~\ref{fig:opt_setup}(c)].
The output from the photodiode was demodulated at two modulation frequencies, $f_{\mathrm{mod, 1}}$ and $f_{\mathrm{mod, 2}} $, to obtain the ODMR signals at each diamond independently using lock-in amplifiers.
The 3 dB cutoff frequency of the low-pass filters in the lock-in amplifiers was 1 kHz.

The center frequencies, $f_{\mathrm{MW, 1}}$ and $f_{\mathrm{MW,2}}$, were continuously stabilized to the resonance frequencies using proportional--integral--derivative (PID) regulators.
The PID bandwidth was set to 200~Hz, which was higher than the encephalomagnetic field frequency range of $<100~\mathrm{Hz}$.
The servo signals, $V_{f, \mathrm{1}}$ and $V_{f, \mathrm{2}}$, sent to the signal generators from the PID regulators shifted the center frequencies by $\alpha_{i}V_{f,i} $, where $\alpha_i$ was a constant that could be set.
As these frequency shifts were equal to the change in the resonance frequencies, the change in the magnetic field $\Delta B_i$ could be expressed as
\begin{equation}
\Delta B_{i}
=\frac{\alpha_i}{\gamma_e} V_{f,i},
\label{eq:deltaB}
\end{equation}
where $\gamma_e$ is the gyromagnetic ratio for the NV centers.

\section{Results and discussion}

\subsection{ODMR}
\begin{figure}
\centering
\includegraphics[scale=1]{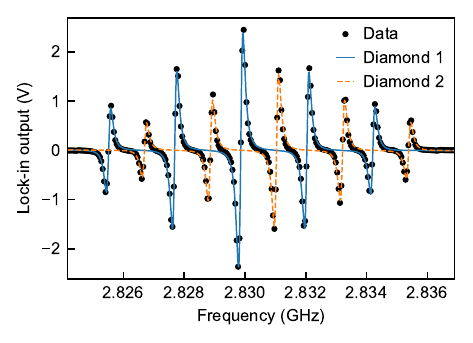}
\caption{\label{fig:lock_in}
    Lock-in ODMR spectrum.
    The blue and orange curves represent the sets of derivative Lorentzian components for diamond 1 and 2, respectively, in the curve fitting.
}
\end{figure}

A representative spectrum of the lock-in ODMR is shown in Fig.~\ref{fig:lock_in}.
In this measurement, we swept the center frequency of a single 3-tone MW field across a resonance frequency of approximately 2.830~GHz.
The strength of the MW field was almost the same as that used in the subsequent measurements.
The modulation frequency was 7.5~kHz.
The observed spectrum was fitted with two sets of summations of derivative Lorentzian functions with a hyperfine splitting of 2.16~MHz.
These sets represented the ODMR spectra of the two diamonds (the blue and orange curves in the figure).
The difference between the bias fields for diamonds 1 and 2 was 42~$\mathrm{\mu T}$.
Given that the linewidths of the peaks were approximately 200 kHz, which were narrower than the hyperfine splitting (2.16 MHz), the magnetic resonance for each diamond was independently used for magnetometry with this relatively small field difference.

\subsection{Calibration}
\label{sec: calib}

\begin{figure}
\centering
\includegraphics[scale=1]{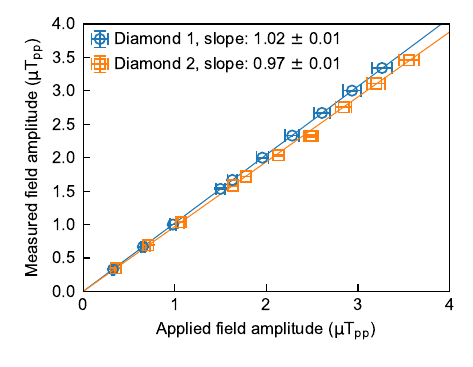}
\caption{\label{fig:calibration}
    Calibration.
    The blue circles and orange squares represent the field amplitudes measured using diamonds 1 and 2, respectively.
    The solid lines are the fitted lines proportional to the applied amplitude.
}
\end{figure}

We calibrated the two-point magnetometry by modulating the current flowing through the coils and measuring the modulation amplitudes in the bias fields.
The current modulation was sinusoidal at 30~Hz, and it generated a sinusoidal component in the applied field for each diamond.
The measured field modulation amplitude as a function of the applied modulation amplitude is shown in Fig.~\ref{fig:calibration}, where the applied amplitude was calculated from the independent calibration of the coils (see Sec.~\ref{sec: setup}) and the modulation amplitude in $I_\mathrm{coil}$.
The blue circles and orange squares represent the field amplitudes measured for diamonds 1 and 2, respectively.
The error bars represent the standard deviation of the mean over multiple measurements.
The uncertainty in the applied field amplitude mainly resulted from the uncertainty in the calibration constants of the coils.
The measured amplitude was fitted using a proportionality function, as indicated by the solid lines, to yield slopes of $1.02\pm 0.01$ and $0.97 \pm 0.01$.
These slopes were close to unity, suggesting that the applied magnetic field was measured successfully using the two-point magnetometry.

We also investigated the bandwidth using diamond 1 by measuring the response to the sinusoidal test field at various frequencies.
The measured amplitude of the test field decreased to $-3~\mathrm{dB}$.
Therefore, we evaluated the 3~dB bandwidth as 300~Hz.
This bandwidth approximately agreed with the bandwidth of the PID feedback control (200~Hz), which limited the measurement bandwidth.

\subsection{Sensitivity}
\begin{figure}
\centering
\includegraphics[scale=1]{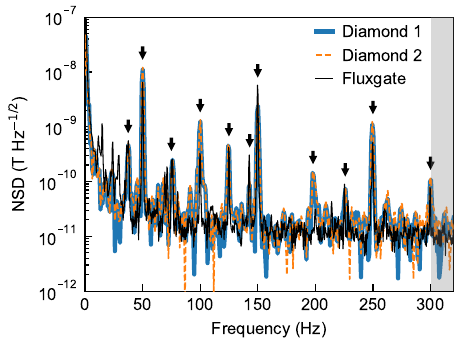}
\caption{\label{fig: onreso}
     Noise spectral density.
     The solid blue and dashed orange lines show the noise spectral density measured using diamonds 1 and 2, respectively, without magnetic shielding.
     The thin black line shows the noise spectral density in the environmental noise measured by a fluxgate sensor adjacent to the diamond magnetometer.
     The arrows indicate the noise peaks that were eliminated from the sensitivity evaluation by applying digital notch filters.
}
\end{figure}

We measured the noise in the developed magnetometer by analyzing the time trace of the output.
This measurement was performed without any magnetic shield.
Figure~\ref{fig: onreso} shows the single-sided noise spectral density obtained from the time trace.
The blue and orange curves represent the noise densities measured using diamonds 1 and 2, respectively.
The two diamonds exhibited a similar noise spectral density with a noise density floor of approximately $15~\mathrm{pT/\sqrt{Hz}}$ at $>100~\mathrm{Hz}$.
We attributed the increase in noise density below 25~Hz to environmental noise. 
This was because the noise density measured on another day at a neighboring location with a fluxgate sensor (black curve) was similar in that frequency region.
Low-frequency environmental noise was produced by trains approximately 20~m away from the setup. 
This noise was not measured by the fluxgate sensor when no trains ran during the period between the last and first trains.
In addition to the low-frequency noise, the agreement between the noise peaks measured using the diamonds and fluxgate sensor (except those at approximately 90~Hz and 200~Hz) implied that most of the noise peaks originated from the environment.
Although the source of the peaks at approximately 90~Hz and 200~Hz was not identified, they were reduced by taking the difference between the outputs for diamonds 1 and 2.
Thus, we considered these peaks to be magnetic noise applied in common mode to diamonds 1 and 2.
We also suspected that they were caused by environmental noise.
We note that the noise floor of the fluxgate sensor was estimated to be approximately $15~\mathrm{pT/\sqrt{Hz}}$ based on the noise density measured inside a magnetic shielding box.

The sensitivity was evaluated using the noise power of the acquired time traces after applying digital filters to suppress the effects of environmental noise.
We applied digital notch filters to the peaks indicated by arrows in Fig.~\ref{fig: onreso} and a digital bandpass filter with 3 dB cutoff frequencies of 25~Hz and 300~Hz.
We defined the sensitivity as $\delta B\sqrt{T}$, which represented the ability to detect a magnetic field $\delta B$ with a signal-to-noise ratio of 1 within measurement time $T$.
According to this definition, the sensitivity was calculated as $\delta B\sqrt{T}=\sigma/\sqrt{2f_{\mathrm{NEP}}}$, where $\sigma$ is the standard deviation and $f_{\mathrm{NEP}}$ is the noise equivalent bandwidth in the filtered time trace.
We obtained sensitivities of $21~\mathrm{pT/\sqrt{Hz}}$ and $22~\mathrm{pT/\sqrt{Hz}}$ for diamonds 1 and 2, respectively.
We also evaluated the sensitivity of the gradiometry by taking the difference between the signals measured by diamonds 1 and 2.
The gradiometry sensitivity was improved to $14~\mathrm{pT/\sqrt{Hz}}$ because the common-mode noise, including environmental noise, was reduced \cite{Zha21, Mas21}.
This was consistent with the environmental noise observed at frequencies below 25~Hz; therefore, we concluded that the dominant noise source was environmental noise. 
The photon shot-noise-limited sensitivities were estimated to be $6.1~\mathrm{pT/\sqrt{Hz}}$ and $7.5~\mathrm{pT/\sqrt{Hz}}$ for diamonds 1 and 2, respectively, based on the detected fluorescence photocurrent and zero-crossing slope of the ODMR peak.

\section{Conclusion}
We developed a DQM that could measure the magnetic field at two points with a spacing of 3.6 mm simultaneously based on frequency-division multiplexing.
Sensitivities of $21~\mathrm{pT/\sqrt{Hz}}$ and $22~\mathrm{pT/\sqrt{Hz}}$ were achieved at the two points in a bandwidth range of 25--300~Hz.
The dominant noise source was considered to be environmental noise. The photon shot-noise-limited sensitivities were estimated as $6.1~\mathrm{pT/\sqrt{Hz}}$ and $7.5~\mathrm{pT/\sqrt{Hz}}$.

The proposed magnetometry method has certain limitations.
The resonance peaks at different positions must be separated by a sufficient distance so that they do not interfere with one another.
Therefore, the minimum spacing between the probed ensembles is limited by the spot size of the excitation laser.
In addition, the geometry of the coils or magnets used to produce different bias fields practically limits the spacing.
Given that the laser spot size can be decreased to the micrometer scale and that micromagnets and microcoils can be used to produce a desired field distribution at the submillimeter scale, realizing multipoint measurement at submillimeter spacing using the proposed method is feasible.
A diamond containing perfectly aligned NV centers \cite{Mic14, Les14, Fuk14} is suitable for increasing the number of measurement points because it shows fewer ODMR peaks, and thus, provides additional room in the frequency space for multiplexing.
The capability of real-time measurement at numerous points with short spacing is beneficial for various applications, including biomagnetic sensing and geophysical research.

\begin{acknowledgments}
This work was supported by the MEXT Quantum Leap Flagship Program (MEXT Q-LEAP) Grant Nos. JPMXS0118067395 and JPMXS0118068379.
\end{acknowledgments}

\bibliography{two_point_measurement}

\end{document}